\documentclass{article}

\usepackage{PRIMEarxiv}

\usepackage[utf8]{inputenc} 
\usepackage[T1]{fontenc}    
\usepackage{hyperref}       
\usepackage{url}            
\usepackage{booktabs}       
\usepackage{amsfonts}       
\usepackage{nicefrac}       
\usepackage{microtype}      
\usepackage{fancyhdr}       
\usepackage{graphicx}       
\usepackage{float}
\usepackage{placeins}
\usepackage{amsmath}
\usepackage{xcolor}
\usepackage{listings}
\usepackage{apacite}

\graphicspath{{media/}}

\definecolor{codebg}{RGB}{240,248,255}
\definecolor{codeframe}{RGB}{180,210,235}

\title{\texttt{MVOS\_HSI}: A Python library for preprocessing agricultural
crop hyperspectral data
}

\author{
  Rishik Aggarwal$^{a}$, Krisha Joshi$^{a}$, Pappu Kumar Yadav$^{a}$,
  Jianwei Qin$^{b}$, Thomas F. Burks$^{c}$, Moon S. Kim$^{b}$ \\[5pt]
  \begin{tabular}{p{4.5cm}p{4.5cm}p{4.5cm}}
    $^{a}$Machine Vision and Optical Sensor (MVOS) Lab, Dept.\ of Agricultural and Biosystems Engineering, South Dakota State University, Brookings, SD 57007, USA &
    $^{b}$USDA/ARS Environmental Microbial and Food Safety Laboratory, Beltsville, MD 20705, USA &
    $^{c}$Department of Agricultural and Biological Engineering, University of Florida, Gainesville, FL 32611, USA \\
  \end{tabular} \\[10pt]
  \thanks{\underline{Corresponding Author:} pappu.yadav@sdstate.edu}
}

\begin{document}
\maketitle

\begin{abstract}
Hyperspectral imaging (HSI) allows researchers to study plant traits non-destructively.
By capturing hundreds of narrow spectral bands per pixel, it reveals details about plant
biochemistry and stress that standard cameras miss. However, processing this data is often
challenging. Many labs still rely on loosely organized collections of lab-specific Matrix Laboratory (MATLAB)
or Python scripts, which makes workflows difficult to share and results difficult to reproduce.
\texttt{MVOS\_HSI} is an open-source Python library that provides an end-to-end workflow
for processing leaf-level HSI data. The software handles everything from calibrating raw
Environment for Visualizing Imagery (ENVI) files to detecting and clipping individual leaves based on multiple vegetation indices
Normalized Difference Vegetation Index (NDVI), Chlorophyll Index - Red Edge (CIRedEdge) and Green Chlorophyll Index (GCI). It also includes tools for data augmentation to create
training-time variations for machine learning and utilities to visualize spectral profiles.
\texttt{MVOS\_HSI} can be used as an importable Python library or run directly from the
command line. The code and documentation are available on GitHub. By consolidating these
common tasks into a single package, \texttt{MVOS\_HSI} helps researchers produce consistent
and reproducible results in plant phenotyping.
\end{abstract}


\begin{table}[h]
\centering
\caption{Code Metadata for \texttt{MVOS\_HSI}}
\begin{tabular}{@{}l p{4cm} p{8cm}@{}}
\toprule
\textbf{Nr.} & \textbf{Code metadata description} & \textbf{Metadata} \\
\midrule
C1 & Current code version & v0.2.1 \\
C2 & Permanent link to code/repository used for this code version &
     \url{https://github.com/MVOSlab-sdstate/mvos_hsi} \\
C3 & Permanent link to Reproducible Capsule & N/A \\
C4 & Legal Code License & MIT License \\
C5 & Code versioning system used & git \\
C6 & Software code languages, tools, and services used &
     Python 3.x; NumPy, SciPy, Matplotlib \\
C7 & Compilation requirements, operating environments \& dependencies &
     Standard scientific Python environment on Windows, Linux, or macOS \\
C8 & If available: Link to developer documentation/manual &
     \url{https://github.com/MVOSlab-sdstate/mvos_hsi} \\
C9 & Support email for questions &
     \href{mailto:pappu.yadav@sdstate.edu}{pappu.yadav@sdstate.edu} \\
\bottomrule
\end{tabular}
\label{tab:metadata}
\end{table}

\section{Availability and Requirements}

\subsection{Availability}

\texttt{MVOS\_HSI} is distributed as a Python package and is intended
to be installable via the Python Package Index (PyPI). The primary development repository,
issue tracker, and user documentation are hosted on GitHub (Github Inc., San Francisco, CA, USA)
(see Table~\ref{tab:metadata}).
The repository includes example commands and recommended dataset organization for
reproducible use.

\subsection{Installation}

\texttt{MVOS\_HSI} is designed to be installed into a dedicated Python (Python
Software Foundation, Wilmington, Delaware, USA) environment
(e.g., a \texttt{venv} or Conda (Anaconda Inc., Austin, TX, USA) environment). For reproducible computational work, we
recommend that users record their exact package versions (e.g., via \texttt{pip freeze}) and
archive configuration files with each analysis, following widely used reproducibility
guidelines~\cite{stodden2014best,sandve2013ten}.

\subsection{Operating Systems}

The library is intended for standard scientific Python environments on
Windows (Microsoft Corporation, Redmond, WA, USA),
Linux (Linux Foundation, San Francisco, CA, USA),
and Macintosh Operating System (macOS) (Apple Inc., Cupertino, CA, USA).
The command-line interface (CLI) and Python Application Programme Interface (API) are platform-independent,
assuming the required dependencies are installed.

\subsection{Dependencies}

\textbf{Core requirements:}~Python 3.x and common scientific Python packages.

\texttt{MVOS\_HSI} is implemented in Python and builds on the scientific Python ecosystem,
relying on Numerical Python (NumPy) for efficient array computing~\cite{harris2020array}, Scientific Python (SciPy) for numerical
routines and scientific utilities, and MATLAB, Plot, and Library (Matplotlib) for visualization~\cite{hunter2007matplotlib}.
The package also integrates with Spectral Python, Scientific Kit (scikit)-image, imgaug, and related libraries
for hyperspectral I/O, calibration, image processing, data augmentation, and diagnostic
plotting.

\subsection{Input and Output Formats}

\textbf{Input:}~The calibration and clipping workflows target raw hyperspectral data stored
in the ENVI format, typically provided as a binary image cube (\texttt{.img}) accompanied by
a header file (\texttt{.hdr}) that stores metadata such as image dimensions, interleave type,
and wavelength information. This format is common in laboratory and field HSI systems, and
the workflow assumes that dark reference imagery is acquired with the same sensor settings
(exposure time, gain, binning) as the sample imagery~\cite{geladi2004hyperspectral}.

\textbf{Output.}~\texttt{MVOS\_HSI} produces:
\begin{enumerate}
  \item Calibrated hyperspectral outputs saved in MATLAB (Mathworks Inc., Natick, MA, USA)
        \texttt{.mat} format (e.g., \texttt{<stem>\_R.mat} and \texttt{<stem>\_F.mat})
        for downstream analysis and compatibility with existing MATLAB-based workflows.
  \item Clipped leaf hypercubes saved as ENVI pairs (\texttt{.hdr/.img}) in a dedicated
        output folder, enabling per-leaf processing and machine-learning dataset preparation.
  \item Augmented hypercubes (ENVI \texttt{.hdr/.img}) generated from clipped leaves using
        configurable augmentation operations.
  \item Spectral plots (on-screen and/or saved figures) for quality control and for
        generating publication-ready spectral profiles.
\end{enumerate}

\section{Motivation and Significance}

Hyperspectral imaging (HSI) has rapidly evolved from a remote-sensing niche into a
fundamental tool for modern plant phenotyping and precision agriculture. Unlike conventional Red, Green, and Blue
(RGB) cameras, which mimic human vision by capturing broad bands of red, green, and blue
light, hyperspectral sensors capture hundreds of contiguous spectral bands across the visible
and near-infrared regions. This results in a three-dimensional hypercube $(x, y, \lambda)$
where every pixel contains a detailed spectral signature. This rich data allows researchers
to non-destructively quantify physiological traits that are invisible to the naked eye, such
as leaf water content, pigment concentration, and early signs of biotic or abiotic
stress~\cite{lowe2017hyperspectral,furbank2011phenomics}. However, the very feature that
makes HSI powerful is its high dimensionality, also makes it difficult to work with. A single
scan of a crop canopy or a set of leaves can generate gigabytes of data, creating a phenotyping
bottleneck where the capacity to generate data far outstrips the ability to process and
interpret it~\cite{tardieu2017plant,thenkabail2018hyperspectral}.

Despite the widespread availability of commercial hyperspectral cameras, the software
ecosystem for processing this data remains fragmented and immature. In many academic and
research settings, the preprocessing workflow i.e taking raw binary data and converting it into
clean, usable information is handled by ad-hoc scripts written in MATLAB or Python. These
scripts are frequently developed by a single student or researcher for a specific experiment
and lack the documentation or structure necessary for long-term maintenance. This kind of
approach leads to significant issues with reproducibility; if a script relies on manual variable
tuning or hard-coded file paths, it becomes nearly impossible for other laboratories (or even
future members of the same lab) to replicate the analysis on new datasets~\cite{stodden2014best}.
Furthermore, common tasks such as removing the background from an image or extracting
individual leaves from a scan are often performed manually using point-and-click Graphical User Interface (GUI) software.
This manual approach is not only tedious and time-consuming but introduces inter-operator
variability that can skew the results of sensitive machine learning models.

\texttt{MVOS\_HSI} was developed to address this specific gap in the research infrastructure.
The motivation for this tool arose directly from our own experience: in recent studies
conducted within our laboratory on hyperspectral detection and severity classification of
Sudden Death Syndrome (SDS) in soybean foliage~\cite{iqbal2025sds,yadav2025sds}, a
significant portion of the effort was devoted to low-level preprocessing tasks extracting
individual leaf reflectance cubes, removing noisy spectral bands, segmenting leaf regions
from background, and managing the resulting large data volumes before any meaningful
machine learning analysis could begin. These recurring preprocessing burdens, which are
common across hyperspectral plant phenotyping studies, motivated the development of a
standardized, reusable solution. We aimed to eliminate the need for researchers to write
``glue code'' to handle basic tasks like parsing ENVI header files, managing dark current
subtraction, or performing complex array slicing for data augmentation. By consolidating
these disparate steps into a standardized, open-source Python library, we provide a robust
foundation for high-throughput phenotyping. This tool allows plant scientists to focus on
the biological questions, such as detecting disease resistance or estimating yield potential,
rather than struggling with the intricacies of multidimensional array manipulation.

\textbf{Objectives.}~The primary design goals of \texttt{MVOS\_HSI} are:
\begin{enumerate}
  \item To encapsulate the entire preprocessing pipeline from raw sensor calibration to the
        generation of machine-learning-ready datasets into a single, installable package.
  \item To enforce a standardized directory structure and workflow, thereby reducing
        configuration errors and ensuring that data is processed consistently across different
        experiments.
  \item To provide a flexible interface that serves both non-programmers (via Command Line
        Interface) and advanced developers (via modular Python API).
  \item To foster reproducibility in the plant science community by providing an open-source
        reference implementation that allows verified, peer-reviewed processing methods to be
        shared alongside research results.
\end{enumerate}

\section{Software Description}

\subsection{Software Architecture}

\texttt{MVOS\_HSI} (\texttt{mvos\_hsi}) is a Python package for hyperspectral preprocessing
in controlled-environment and field phenotyping workflows. The software is designed for two
complementary usage modes: (i)~an importable library for integration into research scripts
and notebooks and (ii)~a command-line interface (CLI) for non-interactive batch processing.
Both interfaces expose the same core capabilities: calibration, leaf clipping, augmentation,
and spectral plotting and operate on folder-structured datasets to support reproducible,
end-to-end processing.

Internally, \texttt{MVOS\_HSI} follows a folder-oriented pipeline. Users specify a dataset
root directory and (when needed) the base path of a dark reference acquisition.
\texttt{MVOS\_HSI} scans for compatible ENVI inputs (paired \texttt{.hdr/.img} files) using
consistent naming conventions, executes the requested processing step(s), and writes derived
outputs to well-defined output folders. This design reduces manual bookkeeping and encourages
standardized dataset organization across experiments and collaborators.

\texttt{MVOS\_HSI} builds on the scientific Python ecosystem. NumPy provides the core
n-dimensional array data model and vectorized operations~\cite{harris2020array}, SciPy
provides MATLAB \texttt{.mat} file interoperability and numerical utilities, and Matplotlib
is used for report-quality visualization~\cite{hunter2007matplotlib}. Hyperspectral I/O
for ENVI datasets is handled through Spectral Python (SPy), enabling reliable loading and
writing of \texttt{.hdr/.img} hypercubes. For common image-analysis operations used in
segmentation (thresholding and connected-component region analysis), \texttt{MVOS\_HSI}
leverages implementations from scikit-image where appropriate~\cite{vanderWalt2014scikit}.
Geometric augmentations are implemented so that each spatial transform is applied consistently
across all wavelength channels, using imgaug-style augmentation primitives~\cite{shorten2019survey}.

\begin{figure}[h!]
  \centering
  {\includegraphics[width=0.8\textwidth]{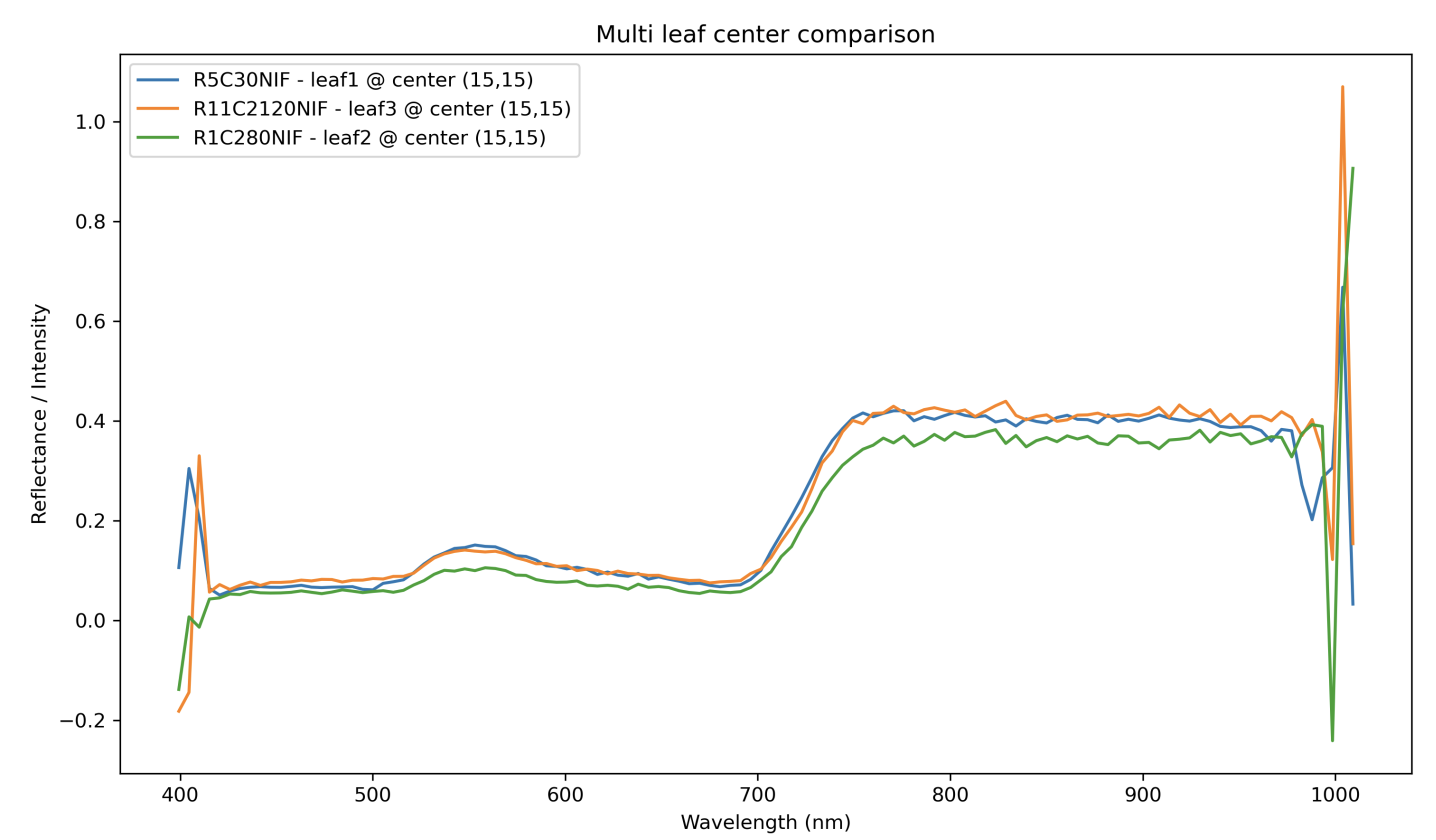}}
  \caption{Example spectral profiles produced by \texttt{MVOS\_HSI} plotting utilities,
  enabling comparison of spectral profiles across multiple leaf samples.}
  \label{fig:spectra}
\end{figure}

The package is organized into modules that mirror the preprocessing pipeline:

\begin{enumerate}
  \item \textbf{Calibration:} Dark-reference correction of raw reflectance-like and
        fluorescence-like measurements with optional spatial and spectral binning.
        The calibration module removes the additive sensor bias (dark current and offset)
        that is present in every raw hyperspectral acquisition. A dark reference cube,
        captured under the same exposure and gain settings as the sample, is subtracted
        pixel-wise and band-wise from the raw data. Where a white reference is unavailable,
        the dark-subtracted output serves as a reflectance-like approximation suitable for
        downstream segmentation and modeling. Spectral binning reduces the number of bands
        by averaging every $k$ adjacent channels, while spatial binning averages over
        $k \times k$ pixel neighborhoods to suppress spatial noise both are
        user-configurable and help manage file size and processing time.

  \item \textbf{Clipping:} Automated detection of leaf regions using vegetation indices
        (NDVI, CIRedEdge, GCI), objective or user-defined thresholding, and cropping into
        per-leaf hypercubes. The clipping module first computes a per-pixel vegetation index
        image from the calibrated hypercube, exploiting the contrast between the spectral
        signatures of plant tissue and background materials. A binary leaf mask is then
        derived either through Otsu's automatic thresholding, which maximizes between-class
        variance, or through a user-supplied threshold when prior knowledge of the scene
        is available. Connected-component analysis is subsequently applied to remove small
        spurious regions caused by sensor noise or debris, and each surviving leaf region
        is independently cropped into its own hyperspectral cube, preserving the full
        spectral information for per-leaf downstream analysis.

  \item \textbf{Augmentation:} Geometry-preserving transformations of clipped leaf
        hypercubes to expand training sets for machine-learning models. Because
        hyperspectral phenotyping datasets are typically small limited by acquisition
        cost, instrument availability, and the time required to scan individual samples
        data-hungry deep learning models are prone to overfitting when trained directly on
        raw collections. The augmentation module addresses this by generating additional
        synthetic training samples through controlled geometric transformations, including
        random rotations, horizontal and vertical flips, and shearing operations. Each
        transformation is applied identically across all wavelength bands of the hypercube,
        ensuring that the spectral signature of every pixel is preserved while its spatial
        context is varied.

  \item \textbf{Plotting:} Spectral visualization utilities for quality control, including
        center-pixel spectra, pixel-specific spectra, ROI mean spectra, and multi-sample
        comparisons. Reliable preprocessing depends on being able to inspect intermediate
        outputs at each stage of the pipeline. The plotting module provides a suite of
        diagnostic tools that allow researchers to visualize the spectral response of
        individual pixels, the mean spectrum of a user-defined region of interest (ROI),
        or overlaid spectral profiles from multiple leaf samples simultaneously. These
        visualizations support both rapid quality-control checks and the generation of
        publication-ready spectral comparison figures, as illustrated in
        Figure~\ref{fig:spectra}.
\end{enumerate}

\begin{figure}[htbp]
  \centering
  {\includegraphics[width=0.5\textwidth]{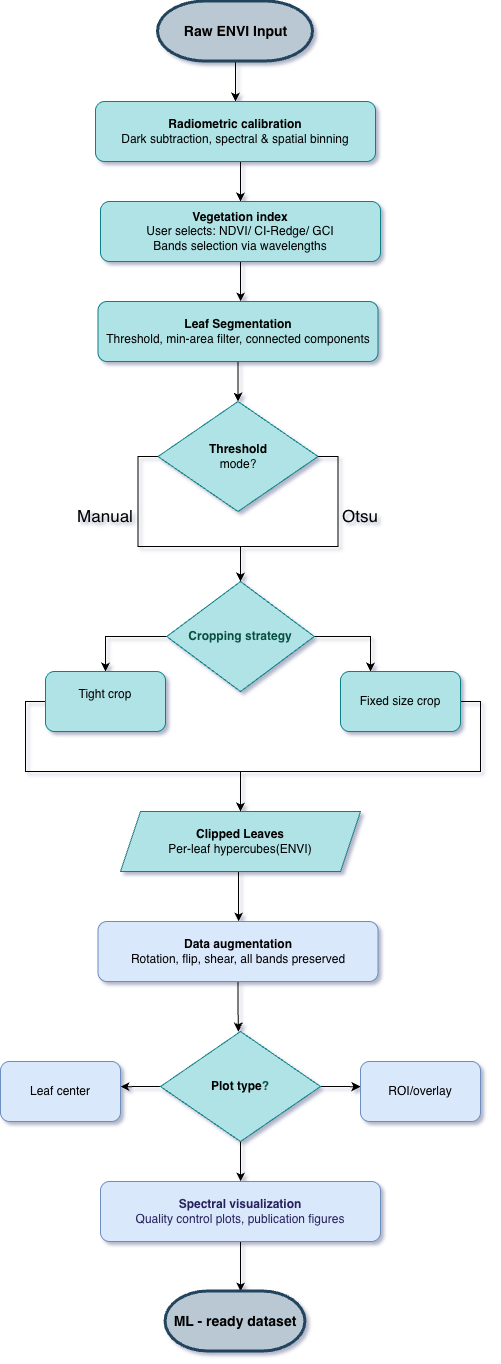}}
  \caption{\texttt{MVOS\_HSI} end-to-end workflow for leaf-level hyperspectral preprocessing.}
  \label{fig:workflow}
\end{figure}

\subsection{Software Functionalities}

This section summarizes the major processing steps implemented in \texttt{MVOS\_HSI} and
clarifies assumptions about expected input data and outputs. The package targets hyperspectral
image cubes stored in ENVI format (\texttt{.hdr/.img}) and assumes that each sample is
acquired alongside a dark reference cube captured under the same sensor configuration
(exposure, gain, binning). In the intended workflow, each sample has paired acquisitions
representing reflectance-like and fluorescence-like measurements (denoted by suffixes
\texttt{\_R} and \texttt{\_F}, respectively), and the corresponding dark reference follows
the same convention (e.g., \texttt{Dark\_R}, \texttt{Dark\_F}). Wavelength information can
be provided through a MATLAB (Mathworks Inc., Natick, MA, USA) file (e.g.,
\texttt{wavelengths.mat} containing a \texttt{wavelength} vector) and/or a one-column CSV
file; if wavelengths are not provided, the software falls back to band-index-based selection
when necessary.

\subsubsection{Radiometric Calibration and Binning}

The calibration step corrects raw sensor measurements using a dark reference cube to remove
the dominant additive bias (dark current and sensor offset), producing a calibrated hypercube
for downstream processing~\cite{geladi2004hyperspectral}. While full reflectance calibration
is often expressed using both dark and white references,

\begin{equation}
  R(\lambda) = \frac{I_{\mathrm{raw}}(\lambda) - I_{\mathrm{dark}}(\lambda)}
                    {I_{\mathrm{white}}(\lambda) - I_{\mathrm{dark}}(\lambda)},
  \label{eq:reflectance}
\end{equation}

\texttt{MVOS\_HSI} is designed to operate robustly in laboratory and greenhouse pipelines
where a white reference may not be recorded for every session. In these cases,
\texttt{MVOS\_HSI} performs dark subtraction to remove the dominant additive component and
produce reflectance-like values suitable for segmentation, clipping, and downstream modeling.

\texttt{MVOS\_HSI} supports two user-controlled binning mechanisms to reduce file size,
suppress noise, and accelerate later stages:

\begin{itemize}
  \item \textbf{Spectral binning} aggregates adjacent wavelength channels by averaging every
        $k$ bands. Users set \texttt{spectral\_bin} to an integer $k \geq 1$ (with $k=1$
        indicating no binning). Invalid settings (e.g., $k=0$) are rejected to avoid ambiguous
        behavior.
  \item \textbf{Spatial binning} averages over local $k \times k$ neighborhoods (user
        parameter \texttt{spatial\_bin}) to reduce spatial noise and stabilize vegetation
        index computations, at the cost of spatial resolution.
\end{itemize}

Calibrated outputs are written in MATLAB-compatible \texttt{.mat} format (e.g.,
\texttt{<stem>\_R.mat} and \texttt{<stem>\_F.mat}), enabling interoperability with existing
MATLAB-based workflows and analysis scripts.

\subsubsection{Vegetation-Index Computation and Leaf Segmentation}

Leaf clipping begins by computing a vegetation index image that emphasizes plant material
relative to background. For example, the normalized difference vegetation index (NDVI) is
\begin{equation}
  \mathrm{NDVI} = \frac{\rho_{\mathrm{NIR}} - \rho_{\mathrm{red}}}
                       {\rho_{\mathrm{NIR}} + \rho_{\mathrm{red}}},
  \label{eq:ndvi}
\end{equation}
where $\rho_{\mathrm{NIR}}$ and $\rho_{\mathrm{red}}$ denote reflectance (or
reflectance-like values) in near-infrared and red bands~\cite{rouse1974monitoring}.
For chlorophyll-sensitive segmentation and visualization, \texttt{MVOS\_HSI} also supports
the green chlorophyll index (GCI),
\begin{equation}
  \mathrm{GCI} = \frac{\rho_{\mathrm{NIR}}}{\rho_{\mathrm{green}}} - 1,
  \label{eq:gci}
\end{equation}
and the red-edge chlorophyll index (CI-RedEdge),
\begin{equation}
  \mathrm{CI}_{\mathrm{RedEdge}} = \frac{\rho_{\mathrm{NIR}}}{\rho_{\mathrm{red\text{-}edge}}} - 1,
  \label{eq:cire}
\end{equation}
where $\rho_{\mathrm{green}}$ and $\rho_{\mathrm{red\text{-}edge}}$ denote reflectance in
the green and red-edge bands respectively~\cite{gitelson2003relationships}. Both indices are
more directly tied to chlorophyll content than NDVI and are useful when the goal is
distinguishing leaves by pigmentation rather than simply separating vegetation from
background.

Given an index image, \texttt{MVOS\_HSI} segments leaf regions using either a user-provided
threshold or an automated threshold selected via Otsu's method~\cite{otsu1979threshold}.
The resulting binary mask is optionally cleaned using simple morphological operations and
filtered by a minimum connected-component area (\texttt{min\_area}) to suppress small false
positives. Each retained region is then cropped from the hyperspectral cube into a per-leaf
hypercube using either fixed-size square crops or tight bounding boxes.

\begin{figure}[h!]
  \centering
  {\includegraphics[width=0.8\textwidth]{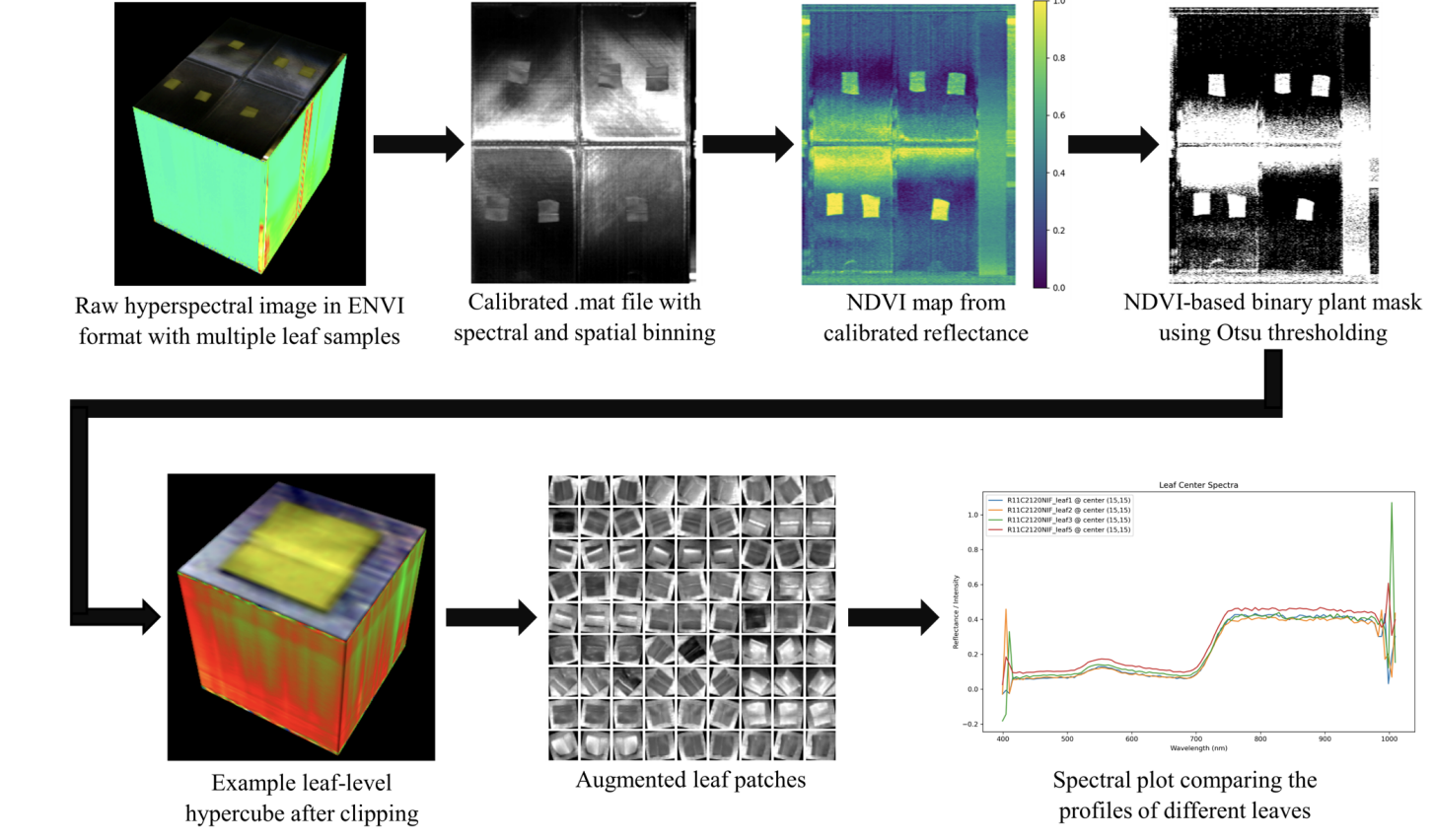}}
  \caption{Example leaf segmentation and clipping workflow using a vegetation index (e.g.,
  NDVI) and objective thresholding (Otsu).}
  \label{fig:segmentation}
\end{figure}

\subsubsection{Hyperspectral Data Augmentation}

Many plant phenotyping datasets are limited by acquisition cost and time, which can lead to
overfitting in data-hungry models. \texttt{MVOS\_HSI} addresses this by applying geometric
augmentations (rotation, flipping, and shearing) to hyperspectral cubes so that each spatial
transform is applied consistently across all wavelength channels~\cite{shorten2019survey}.
This preserves the spectral signature of each pixel while introducing realistic pose
variations that support robust model training.

\subsubsection{Spectral Visualization and Diagnostics}

To help users validate pre-processing choices, \texttt{MVOS\_HSI} provides plotting utilities
to inspect spectra from representative pixels/regions (e.g., the leaf center) and to compare
spectra before and after calibration and clipping.

\section{Illustrative Examples}

A typical project starts with a folder of raw ENVI files and matching dark-reference files
in the same folder.

\paragraph{1) Setup and path configuration}

\begin{lstlisting}[caption={Pipeline setup}, label={lst:setup}]
from pathlib import Path
import mvos_hsi

root = Path(r"C:\path\to\your\dataset")
dark_base = root / "Dark"
wavelengths_mat = Path(r"C:\path\to\wavelengths.mat")
clips_outdir = root / "clipped_hypercubes"
\end{lstlisting}

This setup block defines the dataset root, dark-reference files, wavelength metadata file,
and expected clipping output directory used by subsequent steps.

For command-line use, \texttt{mvos-hsi} is available as a global command after installation:

\begin{lstlisting}[language=bash, caption={CLI setup and help}, label={lst:cli-help}]
mvos-hsi --help
\end{lstlisting}

\noindent
On Windows, use \texttt{\^{}} for line continuation; on Linux/macOS, use \texttt{\textbackslash}.

\paragraph{2) Calibration}

\begin{lstlisting}[caption={Calibrate raw ENVI data}, label={lst:calibrate}]
mvos_hsi.calibrate_folder(
    folder=str(root),
    dark_base=str(dark_base),
    spectral_bin=3,
    spatial_bin=3,
)
\end{lstlisting}

\texttt{calibrate\_folder} applies dark correction and binning to raw hyperspectral cubes,
producing calibrated outputs suitable for segmentation and analysis.

The same stage can also be run from the command line:

\begin{lstlisting}[language=bash, caption={CLI example for calibration}, label={lst:cli-calibrate}]
mvos-hsi calibration folder \
  --folder "C:\path\to\dataset" \
  --dark   "C:\path\to\dataset\Dark" \
  --k 3 \
  --spatial 3
\end{lstlisting}

\paragraph{3) Leaf clipping and augmentation}

\begin{lstlisting}[caption={Clip leaves and generate augmented samples}, label={lst:clip}]
clip_result = mvos_hsi.clip_folder(
    folder=str(root),
    index="ndvi",
    wavelengths_mat=str(wavelengths_mat),
    threshold_mode="auto",  # Otsu thresholding
    min_area=100,
    crop_mode="square",
    crop_size=30,
)

mvos_hsi.augment_folder(
    folder=str(clips_outdir),
    num_aug=3,
    flip=True,
    rotate=(-10, 10),
    shear=(-16, 16),
)
\end{lstlisting}

In this stage, \texttt{clip\_folder} isolates leaves from background and writes per-leaf
cubes; then \texttt{augment\_folder} expands the dataset with geometry-preserving transforms
for machine learning training.

Equivalent CLI commands are shown below for clipping and augmentation:

\begin{lstlisting}[language=bash, caption={CLI example for clipping}, label={lst:cli-clipping}]
mvos-hsi clipping folder \
  --folder          "C:\path\to\dataset" \
  --index           ndvi \
  --wavelengths-mat "C:\path\to\wavelengths.mat" \
  --threshold-mode  auto \
  --crop-mode       square \
  --crop-size       30
\end{lstlisting}

\begin{lstlisting}[language=bash, caption={CLI example for augmentation}, label={lst:cli-augmentation}]
mvos-hsi augmentation folder \
  --folder "C:\path\to\dataset\clipped_hypercubes" \
  --num 3 \
  --flip \
  --rotate -10 10 \
  --shear  -16 16
\end{lstlisting}

\paragraph{4) Spectral diagnostics}

\begin{lstlisting}[caption={Plot representative leaf spectra}, label={lst:plot}]
mvos_hsi.plot_leaf_center(
    clipped_dir=str(clips_outdir),
    stem="H_P1_V4_B",
    leaves=[1, 2],
    wavelengths_mat=str(wavelengths_mat),
    title="Center pixel spectra",
    show=True,
)
\end{lstlisting}

\texttt{plot\_leaf\_center} provides a quick quality-control view of spectral behavior from
selected leaves before statistical or machine-learning analysis.

The plotting utilities are also accessible from the CLI:

\begin{lstlisting}[language=bash, caption={CLI example for single-sample spectral plotting}, label={lst:cli-plot-single}]
mvos-hsi plotting leaf \
  --clipped-dir     "C:\path\to\clipped_hypercubes" \
  --stem            H_P1_V4_B \
  --leaf            1 3 \
  --wavelengths-mat "C:\path\to\wavelengths.mat"
\end{lstlisting}

\begin{lstlisting}[language=bash, caption={CLI example for multi-sample spectral comparison}, label={lst:cli-plot-multi}]
mvos-hsi plotting leaf-multi \
  --clipped-dir     "C:\path\to\clipped_hypercubes" \
  --item            H_P1_V4_B:1 \
  --item            H_P1_V6_B:3 \
  --wavelengths-mat "C:\path\to\wavelengths.mat"
\end{lstlisting}

\section{Impact}

The main goal of \texttt{MVOS\_HSI} is to make hyperspectral analysis more accessible for
plant-science researchers who need to convert raw sensor outputs into analysis-ready datasets.
By packaging common preprocessing steps radiometric correction, segmentation, cropping,
augmentation, and visualization into a single library, \texttt{MVOS\_HSI} reduces the
amount of custom code that individual labs need to write and maintain, and encourages more
consistent preprocessing decisions across studies.

\begin{itemize}
  \item \textbf{Reproducibility and transparency:} Many existing hyperspectral workflows
        rely on manual GUI-based tools that are difficult to document and hard for others to
        replicate exactly. Scripted pipelines make every preprocessing decision explicit,
        which makes it easier for collaborators and reviewers to verify or reproduce a given
        analysis~\cite{stodden2014best,sandve2013ten,wilson2014best}.

  \item \textbf{Reduced phenotyping bottlenecks:} In plant phenotyping, the rate at which
        data can be collected has outpaced the capacity to analyze it, and manual processing
        steps are a large part of why~\cite{tardieu2017plant}. Automated segmentation and
        batch processing help close that gap and make higher-throughput workflows more
        practical for everyday lab use.

  \item \textbf{Machine-learning readiness:} Labeled hyperspectral datasets tend to be small
        because collecting and annotating them is time-consuming and expensive. The built-in
        augmentation utilities give users a straightforward way to generate controlled
        training-time variability, which is a well-established approach for improving model
        performance under data-scarce conditions~\cite{shorten2019survey}.

  \item \textbf{Interoperability:} Outputs from \texttt{MVOS\_HSI} follow standard array
        and dataframe conventions, so they work directly with common Python tools like
        scikit-learn and TensorFlow without needing additional format conversion. This makes
        it easier to slot the library into existing workflows rather than building around it.
\end{itemize}

\section{Conclusions}

\texttt{MVOS\_HSI} provides an open-source, end-to-end workflow for preprocessing leaf-level
hyperspectral data, covering calibration, automated leaf segmentation and clipping,
augmentation, and visualization. Before tools like this existed, researchers typically had
to either rely on proprietary software with limited scripting support or piece together
their own pipelines from scratch both of which make it harder to share methods, catch
errors, and reproduce results across studies. By consolidating these operations into a single
package with both a Python API and a CLI, \texttt{MVOS\_HSI} reduces that fragmentation and
gives labs a common starting point for hyperspectral preprocessing. In practical terms, the
library is designed to help researchers move from raw ENVI files to analysis-ready outputs
with fewer manual steps and more transparency about how the data were processed. The
calibration and segmentation steps are parameterized, so the exact choices made during
preprocessing are recorded rather than buried in a series of GUI clicks that are difficult
to reconstruct later. The augmentation utilities make it easier to prepare small
hyperspectral datasets for machine-learning tasks, where data scarcity is a common obstacle.
Taken together, these features are intended to lower the barrier for plant-science teams
that want to adopt hyperspectral imaging but do not have the engineering resources to build
and maintain a full preprocessing stack internally.

\section{Limitations and Future Work}

Like any hyperspectral preprocessing workflow, the quality of \texttt{MVOS\_HSI}'s outputs
depends heavily on acquisition conditions. Calibration assumes reasonably stable illumination
and well-collected reference measurements; when those conditions are not met, downstream
results can degrade. Similarly, the current segmentation approach works well for
controlled-environment setups but may struggle with highly variable backgrounds, overlapping
leaves, severe shadowing, or sensor-specific artifacts that the pipeline does not yet
explicitly account for. There are also limitations in terms of sensor coverage. The current
implementation is designed around a specific set of ENVI-compatible formats, and data
collected with sensors that use different file structures or spectral configurations may
require manual adaptation before the pipeline can be applied directly. For future
development, we plan to expand the range of supported sensor formats and reader backends,
add more vegetation index options suited to difficult or variable scenes, and improve
parameter auto-tuning so the pipeline requires less manual calibration when applied to new
datasets.

\section*{Acknowledgements}

The authors would like to thank members of the Machine Vision and Optical Sensor (MVOS) lab
at South Dakota State University for their feedback on the design and testing of
\texttt{MVOS\_HSI} Python library. We also thank Naveen Duhan for his valuable suggestions
regarding the command line interface (CLI).


\bibliographystyle{apacite}
\bibliography{references}

\end{document}